# Evaluation of Motor Imagery-Based BCI methods in neurorehabilitation of Parkinson's Disease patients

A. Miladinović, M. Ajčević, P. Busan, J. Jarmolowska, G. Silveri, Member, IEEE, M. Deodato, S. Mezzarobba, P. P. Battaglini, A. Accardo

*Abstract*— The study reports the performance of Parkinson's disease (PD) patients to operate Motor-Imagery based Brain-Computer Interface (MI-BCI) and compares three selected pre-processing and classification approaches. The experiment was conducted on 7 PD patients who performed a total of 14 MI-BCI sessions targeting lower extremities. EEG was recorded during the initial calibration phase of each session, and the specific BCI models were produced by using Spectrally weighted Common Spatial Patterns (SpecCSP), Source Power Comodulation (SPoC) and Filter-Bank Common Spatial Patterns (FBCSP) methods. The results showed that FBCSP outperformed SPoC in terms of accuracy, and both SPoC and SpecCSP in terms of the false-positive ratio. The study also demonstrates that PD patients were capable of operating MI-BCI, although with lower accuracy.

*Clinical Relevance*— The study presents evidence on how well PD's patients are able to perform Motor-Imagery BCI based neurorehabilitation and reports a comparison of classification accuracy for three selected approaches.

## I. INTRODUCTION

BCI neurorehabilitation, also known as Neurofeedback, refers to the closed-loop utilization of real-time acquisition of neural data that's then transformed and prepared for the extraction of relevant features. The final outcome of machine-learning is then presented back to the subject in the form of visual, auditory, or tactile feedback. Hence, with practice, reinforcement, and feedback, subjects can learn to volitionally control neural activity that has been shown to positively affect cognitive capabilities, motor execution, and coordination, in healthy individuals, as well as in patients, such as post-Stroke and Parkinson's disease [1]-[2], Autism spectrum disorders [3] etc. The most common motor symptoms in Parkinson's Disease (PD) are tremors, rigidity and gait disorders, such as freezing of gait (FOG) and festination [4]. In particular, the literature [1], [5] reports that Motor-Imagery (MI) based BCI (MI-BCI) results in activation of the visual, motor and premotor cortex and as a consequence in an improvement in the individual's locomotor ability, and reduction of the PD symptoms, such as bradykinesia, FOG episodes and rigidity [6]. Besides, real-time feedback also allows a more controlled rehabilitation process since it reveals directly whether the patient performs the given task correctly. Common Spatial Pattern (CSP) filters [7] are one of the most used approaches in the BCI domain, particularly in the context of the MI oscillatory paradigm. This data-driven approach assigns weights to each channel, and it is designed to maximize and minimize the variance for the MI task and rest, respectively. In the past decade different extension of the basic CSP has been proposed, and most commonly used are Filter Bank CSP (FBCSP) [8], in which a series of CSP filters are implemented for different frequency subbands creating frequency-specific task-related model, Spectrally Weighted CSP (SpecCSP) [9] that exploits interactions between frequency bands by assigning a weight to each frequency band, and finally, Source Power Co-Modulation (SpoC) [10] in which the variance is maximized on the component space, instead of on the raw EEG sensor space, as in the case of previous. The average reported accuracy of mention approaches exceeds 70% [11]–[13] and, in some cases, reaches 85% [14]. The most significant disadvantage is that most of the approaches have been tested on healthy individuals, whereas the tests on clinical populations, similar to the one reported in [15], are quite rare. It is not to neglect that apart from BCI illiteracy present in healthy individuals, the clinical population is additionally characterized by cognitive decline [2], especially evident in the domain of executive functions, and therefore, may present different BCI performance.

This study aims to investigate the performance BCI approaches on Parkinson's disease patients and to report a comparison of three selected approaches.

## II. MATERIALS AND METHODS

### A. Study population

The experiment was conducted on 7 patients (4 males and 3 females) with a mean age of 72 years old (standard deviation = 4.5). All patients had a history of gait's disturbance, namely experiencing freezing of gate episodes (FOG), Hoehn and Yahr score lower than 3, whereas, the cognitive capabilities were evaluated by the Mini-Mental State Examination (MMSE). Moreover, all of them had a stable pharmacological treatment for at least two months prior to the neurofeedback treatment.

The recruited patients gave their signed consent before the start of treatment, and the experimental protocol was pre-approved by the Local Ethical Committee and was conducted according to the principles of the Declaration of Helsinki.

* A. Miladinović is supported by the European Social Fund (ESF) - FVG. This work is partially supported by Master's in Clinical Engineering - University of Trieste.

A.Miladinović, M.Ajčević, G.Silveri and A.Accardo are members of the Department of Engineering and Architecture at the University of Trieste, Trieste, Italy (corresponding author phone: +390405587130; e-mail: aleksandar.miladinovic@phd.units.it last author e-mail: accardo@units.it).

P.Busan, J.Jarmolowska and P.P.Battaglini are members of the Department of Life Sciences at the University of Trieste, Trieste, Italy

M, Deodato and S.Mezzarobba are members of Department of Medicine, Surgery and Health Sciences at the University of Trieste, Trieste, Italy

### B. BCI protocol

The BCI protocol consisted of a total of 14 neurofeedback sessions targeting lower extremities with a duration of 1.5-2 hours each repeated 2-3 times per week. The session was split into two parts, the initial calibration phase where the patients had to perform feet MI on a given written instruction "start" shown on the pc monitor for 35 to 40 times, and the online phase where they had to control the stimulus on the screen (feedback) actively. In order to investigate the performance of BCI approaches, this study focused on calibration phase dataset. The EEG signals were acquired from 11 electrodes placed at standard 10-20 locations (F3, Fz, F4, T3, C3, Cz, C4, T4, P3, Pz, P4). All electrodes were referenced to AFz and grounded to POz and the acquisition has been performed with a sampling frequency rate of 256 Hz and impedances were kept below 5kΩ. In addition, two electromyography electrodes were added to exclude any possible limb movement.

### C. EEG processing

The processing of EEG data was carried out using MATLAB (The MathWorks Inc., Natick, MA). All channels were filtered from 6 to 32 Hz with the 2nd order Butterworth bandpass filter and resampled to 128Hz. The BCI models were produced with the BCILAB [16] framework applying three selected BCI classification approaches.

### D. BCI Approaches

The selection of the approaches was based on their performance on healthy individuals. In the case of SpecCSP the reported accuracy varies from 70-80% [13], [17], for the SpoC 76% [12], and up to 90% [11]–[13] in the case of FBCSP. SpecCSP is an advanced paradigm for oscillatory processes using the spectrally weighted CSP algorithm. The approach is applicable to most oscillatory processes and generally gives better results than a CSP with a suitably unrestricted spectral filter (e.g. wideband). As in the case of FBCSP, it exploits interactions between frequency bands buy assigning a weight to each frequency band. The SpoC (Source Power Comodulation) is the youngest among the tested approaches. The reported advantage of this approach is that the features are not extracted directly from the raw time-series EEG data, but it works on component-based, such as ICA, or beamforming space. It is comparable with CSP since, also in this case, the spatial features are learned using the same optimization algorithm. The biggest advantage of this approach is that it optimizes component space, instead of a raw EEG, and thus it is less affected by non-task related oscillatory activities or other external noises. At the same time, it might be prone to artifacts, such as voluntary and involuntary muscular activity, that can co-occur during task performance, especially in the elderly and/or clinical populations. Such a signal has orders of magnitude higher amplitude that variance optimization algorithm might misinterpret as a task-related signal. Finally, the FBCSP [8] is an extension of the basic CSP method, in which a series of CSP filters are implemented for different frequency subbands creating specific frequency and task-related models. The algorithm shows the best performance on the tasks when the oscillatory processes are present in different frequency bands and on different spatial locations. It is designed to provide the best results in the case of complex EEG dynamics and non-trivial interactions between frequency bands, such as the mu/beta ratio in the case of motor imagery task.

For each band, log-variance (power) features are extracted and concatenated and fed to the Fisher's LDA classifier with automatic shrinkage parameter estimation [18].

Finally, the classification accuracy was estimated using 10-fold chronological/blockwise cross-validation with 5 trials margin. Apart from accuracy, type I and type II error parameters were extracted for each session. During the approach selection process we included only algorithms that do not require specialized hardware (such as computer clusters or GPU), nor tedious tune-up of various parameters; therefore, the model can be produced on standard portable computers, in a reasonable time of 5-10min, allowing equipment mobility and applicability in different environments.

### E. Statistical analysis

Differences in classification accuracy and related model performance parameters (True positive ration - TP; False Positives - FP, TPR- True positives ratio; TNR - True negatives ratio; FPR - False positives ratio; FN - False negatives ratio) among evaluated approaches were tested by repeated-measure analysis of variance (ANOVA). Bonferroni corrections were used for post-hoc multiple comparisons.

## III. RESULTS

Classification accuracy obtained by models produced by SpecCSP, SpoC and FBCSP methods is reported in Table 1 for each of the 7 subjects observed in over 14 BCI sessions. Classification accuracy resulted significantly higher for FBCSP (65.2±11.3) and SpoC (63.4±10.3) compared to SpecCSP (60.7±11.5) (p-value <0.001 and 0.015, respectively). No significant difference in total accuracy was found between FBCSP and SpoC (p-value 0.219), although FBCSP presented a slightly higher overall average accuracy and resulted in the lowest error in 5 of 7 PD subjects.

Table 2 reports mean ± SD values of TNR TPR, FPR and FNR observed on our sample for each of the three methods applied. FPR was significantly lower and TNR was significantly higher for FBCSP than those observed in SpecCSP (p-value <0.046) and SpoC (p-value 0.015). TPR was significantly higher and FNR was significantly lower for FBCSP and SpecCSP compared to SpoC (p-value <0.001 and =0.001, respectively).

## IV. DISCUSSION

Motor-Imagery BCI can improve locomotor ability and alleviate some symptoms in PD patients. The ability of the BCI-naïve Parkinson's patients to use BCI based Motor-Imagery neurorehabilitation and choice of appropriate classification method is still debated. This study investigated the performance of these subjects. to use this advanced neurorehabilitation strategy, and furthermore, which among selected approaches is more appropriate for the aforementioned population. In particular, we tested SpecCSP, SpoC and FBCSP on 7 Parkinson's disease patients

TABLE I. CLASSIFICATION ACCURACY (%) OBTAINED BY THE SPECCSP, SPOC AND FBCSP METHODS, RESPECTIVELY, FOR 7 PD PATIENTS OVER 14 BCI SESSIONS PERFORMED FOR EACH SUBJECT.

|  | Approaches | | |
| --- | --- | --- | --- |
| Subjects | SpecCSP | SpoC | FBCSP |
| 1 | 50.4 ± 9.1 | 54.1±6.8 | **62.4±10.3** |
| 2 | 50.9±7.7 | **55.1±6.9** | 53.3±10.3 |
| 3 | 63.7±6.8 | **67.1±4.6** | 64.5±8.9 |
| 4 | 57.6±8.2 | 62.7±7.2 | **70.0±6.1** |
| 5 | 59.1±11.3 | 63.5±4.4 | **65.6±11.2** |
| 6 | 58.2±7.2 | 58.4±7.1 | **63.9± 7.4** |
| 7 | 73.9±13.7 | 74.9±11.6 | **75.2±10.6** |
| Average | 60.7±11.5 | 63.4 ± 10.4 | 65.2±11.3 |

For each patient, the highest accuracy is marked in boldface.

performing feet MI task over 14 sessions.

The main finding of the study is that FBCSP outperforms SpoC and also on average, gives better results than the SpecCSP.

The reason for low SpoC performance might be explained with the fact that the algorithm maximizes components subspace, and since in PD's, noises (i.e. muscular activity due to tremors) can co-occur with the task, the approach misinterprets the noise with the signal. Regarding SpecCSP and FBCSP, there are no significant differences among them, and both approaches seem to be more appropriate when the task elicits changes in power band ratios at the particular scalp locations, such as mu/alpha power decrease and beta increase (beta rebound) [19].

In addition to accuracy, we have also investigated type I and type II errors. The robustness of the BCI approach is reflected by False positives (type I error), which is defined as the existence of relative feedback without a subject's participation which gives the impression that the system is not working, reducing the motivation for participation. Also, in this case, the FBCSP demonstrates superior results.

The study also demonstrated that PD patients were capable of operating MI-BCI, although with lower accuracy. A possible explanation is the existence of possible EEG alternation due to condition progression and medical treatment [20]. Furthermore, a possible cause of their lower performance might be the cognitive decline that is one of the most comorbidity in PD's. A high accuracy method, such as FBCSP, may be used as a tool to instruct subjects to properly perform MI in the initial phases of the standard physiotherapeutic procedures. The results obtained and clinical efficacy of this type of rehabilitation should be confirmed in a larger clinical study.

TABLE II. COMPARING THE PERFORMANCE OF THE SPECCSP, SPOC AND FBCSP METHODS. TPR- TRUE POSITIVES RATIO; TNR - TRUE NEGATIVES RATIO; FPR - FALSE POSITIVES RATIO; FN - FALSE NEGATIVES RATIO

|  | Approaches | | |
| --- | --- | --- | --- |
|  | SpecCSP | SpoC | FBCSP |
| TPR (%) | 59.9±13.9 | 64.6±12.8 | 65.2±12.9 |
| TNR (%) | 61.4±12.1 | 61.9±12.5 | 65.0±13.0 |
| FPR (%) | 38.6±12.1 | 38.1±12.5 | 35.0±13.0 |
| FNR (%) | 40.1±13.9 | 35.4±12.8 | 34.8±12.9 |

## V. CONCLUSION

In conclusion, this preliminary study showed that among selected approaches FBCSP provides the best performance in PD subjects and that they are able to perform BCI based MI neurorehabilitation with relatively high accuracy.